\documentclass[aps,prl,twocolumn,groupedaddress,showpacs]{revtex4}
\usepackage{graphicx}
\usepackage{wasysym}
\begin{document}
\title{Quantum Hall Effect of Dirac Fermions in Graphene: Disorder
Effect and Phase Diagram}
\author{D. N. Sheng$^{1}$, L. Sheng$^{2}$ and Z. Y. Weng$^3$}
\address{
$^1$Department of Physics and Astronomy, California State
University, Northridge, California 91330\\
$^2$Department of Physics and Texas Center for Superconductivity,
University of Houston, Houston, Texas 77204\\
$^3$Center for Advanced Study, Tsinghua University, Beijing
100084, China}
\begin{abstract}
We numerically study the interplay of band structure, topological
invariant and disorder effect in two-dimensional electron system
of graphene in a magnetic field. Two \emph{distinct} quantum Hall
effect (QHE) regimes exist in the energy band with the
unconventional \textquotedblleft half-integer\textquotedblright\
QHE appearing near the band center, consistent with the
experimental observation. The latter is more robust against
disorder scattering than the conventional QHE states near the band
edges. The phase diagram for the unconventional QHE is obtained
where the destruction of the Hall plateaus at strong disorder is
through the float-up of extended levels toward band center and
higher plateaus always disappear first. We further predict a new
insulating phase between $\nu =\pm 2$ QHE states at the band
center, which may explain the experimentally observed resistance
discontinuity near zero gate voltage.
\end{abstract}
\pacs{73.43.Cd; 73.40.Hm; 72.10.-d; 72.15.Rn}
\maketitle
With the advances in micromechanical extraction and fabrication
techniques, high mobility single atomic layer of graphite, called
graphene, became available recently~\cite{G0,G1,G2,G3}, which has
attracted much experimental and theoretical interest. This new
material has many extraordinary properties such as submicron-scale
ballistic transport at room temperature, ability to sustain high
electric currents, and ease of tuning carrier density for both
electrons and holes. It is generally believed that graphene may
have potential applications in electronic
devices~\cite{G0,G1,G2,G3}. Graphene is of great fundamental
interest as well because of its special band structure~\cite{S0}. Undoped
graphene has one $\pi$ electron per lattice site forming a
two-dimensional (2D) extended electronic structure. The only
states at the Fermi energy ($E_{F}=0$) are at two corners of the
Brillouin zone, where the conduction and valence bands touch. The
low energy excitations have a linear dispersion relation similar
to that of the massless Dirac equation, where the electron Fermi
velocity plays the role of an effective ``speed of light''. So the
electron system of graphene can be viewed as a condensed-matter
realization of relativistic Dirac fermions, and the band touching
points are often referred to as Dirac points.

Remarkably, novel quantum Hall effect (QHE) with Hall plateaus
obeying the unconventional quantization rule
$\sigma_{xy}=(k+1/2)g_s \frac {e^2} h$ has recently been observed
experimentally~\cite{Hall0,Hall1} in graphene films in a strong
magnetic field. Here, $k$ is an integer and $g_s=4$ stands for the
spin and sublattice-related degeneracy. In units of
$g_{s}\frac{e^{2}}{h}$, the \textquotedblleft
half-integer\textquotedblright\ quantization of $\sigma _{xy}$ was
conjectured to be due to the nontrivial Berry phase of the Dirac
fermions in a magnetic field~\cite{Hall0,Hall1,phystoday}.
Interestingly, calculations~\cite{T1,T2,T3,T0} by using an analogy
to the $2+1$ dimensional Quantum Electro Dynamics have predicted a
``half-integer'' quantized QHE for graphene. Disorder effect has
been studied by using self-consistent Born approximation within
the continuous model~\cite{T0}. However, so far the interplay of
the band structure, particle-hole symmetry, disorder effect, and
topological property of the energy band in the unconventional QHE
has not yet been investigated. It is thus highly desirable to
perform exact numerical calculations by taking into account the
full band structure and random disorder, in order to reveal the
fundamental nature of the new QHE phases and related quantum phase
transitions in graphene.

In this Letter, we study the QHE in graphene using a tight-binding
model. The experimentally observed unconventional Hall plateaus
are reproduced around the band center $E_{F}=0$, while the
conventional integer QHE plateaus appear near the band edges. The
unusual distribution of the topological invariant quantity-Chern
number in the energy band and the conservation of total Chern
number (which is also the geometric Berry phase~\cite{thouless1})
for the particle and hole bands account for the unconventional
\textquotedblleft half-integer\textquotedblright\ QHE. The latter
is found to be much more stable than the conventional ones near
the band edges against disorder scattering. We map out the whole
phase diagram for the QHE and demonstrate that the QHE plateaus
are destroyed at strong disorder (or weak magnetic field) through
the float-up of extended levels toward the band center. We further
identify a new insulating phase between $\nu=\pm 2$ QHE states,
which may explain the experimentally observed discontinuity in
resistance\cite{Hall0,Hall1} near zero gate voltage.

We consider a rectangular sample of 2D graphene sheet consisting
of carbon atoms on a honeycomb lattice~\cite{S0,lisheng}, which
has totally $L_y$ zigzag chains with $L_{x}$ atomic sites on each
chain~\cite{lisheng}. The size of the sample will be denoted as
$L_{x}\times L_{y}$. In the presence of an applied magnetic field
perpendicular to the graphene plane, the lattice model can be
written in the tight-binding form~\cite{S0,sheng}:
\begin{eqnarray}
H&=&-t\sum\limits_{\langle
ij\rangle\sigma}e^{ia_{ij}}c_{i\sigma}^\dagger
c_{j\sigma}+\sum\limits_{i}w_ic_{i\sigma}^\dagger c_{i\sigma}\ ,
\label{HAMIL}
\end{eqnarray}
where $c_{i\sigma}^+ $ ($c_{i\sigma}$) creates (annihilates) a
$\pi$ electron of spin $\sigma$ on lattice site $i$ 
with $t$ as the nearest-neighbor hopping integral, 
and $w_i$ is random disorder potential uniformly distributed in the interval
$w_i\in [-W/2, W/2]$. The magnetic flux per hexagon
$\phi=\sum_{{\small {\mbox{\hexagon}}}}a_{ij}=\frac{2\pi}{M}$ is
proportional to the strength of the applied magnetic field $B$.
This tight-binding model is valid for describing the full energy
band and realizes the energy dispersion relation of the Dirac
fermions near the band center~\cite {S0}.
\begin{figure}[tbp]
\includegraphics[width=3.2in]{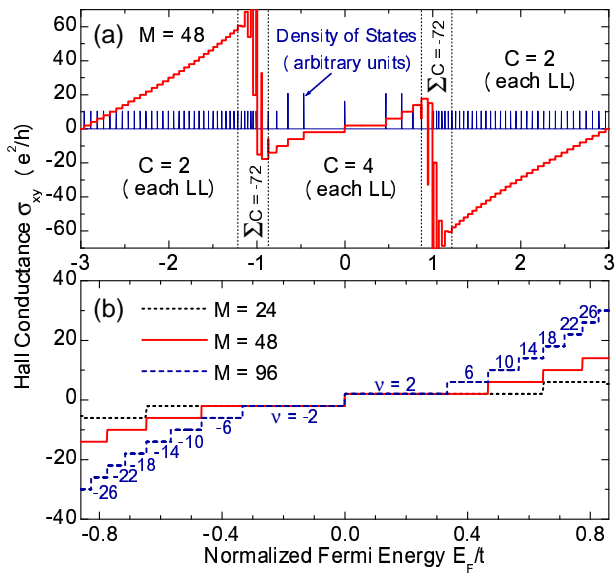}
\caption{(a) Calculated Hall conductance and electron density
of states in the full energy band for magnetic flux strength $%
\protect\phi=\frac{2\protect\pi}{48}$ or $M=48$, and (b) the
unconventional Hall conductance for three different strengths of
magnetic flux near the band center. Here the disorder strength is
set to $W=0$ and the sample size is taken to be $96\times 48$.}
\vskip-0.25cm
\end{figure}

The Hall conductance $\sigma_{xy}$ can be calculated by using the
Kubo formula through exact diagonalization of the system
Hamiltonian~\cite{sheng}. In Fig.\ 1a, the Hall conductance
$\sigma_{xy}$ and electron density of states are plotted as
functions of electron Fermi energy $E_{F}$ for a clean sample
$W=0$ with magnetic flux $\phi=\frac{2\pi}{48}$, which illustrates
the overall picture of the QHE in the full energy band. According
to the behavior of $\sigma_{xy}$, the  energy band is naturally
divided into three different regimes. Around the band center,
$\sigma_{xy}=\nu \frac{e^2}{h}$ is indeed quantized according to
the unconventional quantization rule $\nu=(k+1/2)g_s$ with a
degeneracy factor $g_{s}=4$ for each Landau level (LL) due to two spin
directions and two Dirac points. These Hall plateaus explain the
experimentally observed unconventional QHE~\cite{Hall0,Hall1} and
agree with the results from the theory based upon the continuous
model~\cite{T1,T2,T3,T0}. In Fig.\ 1b, the Hall conductance in
this unconventional region for three different strengths of
magnetic flux is shown. With decreasing magnetic flux from
$\phi=\frac{2\pi}{24}$ to $\phi=\frac{2\pi}{96}$, more quantized
Hall plateaus emerge following the same quantization rule. We see
that the widths of the plateaus are roughly proportional to
$\sqrt{\phi}\propto \sqrt{B}$, in agreement with the continuous
theory~\cite{T1,T2}. The unconventional QHE in the present band
model can be understood in terms of the topological invariant
Chern number~\cite{sheng,bhatt}. Inside each LL, there are
extended states characterized by a nonzero Chern integer. The
total Hall conductance in units of $\frac{e^2}{h}$ is exactly the
sum of the Chern numbers of all the occupied extended
states~\cite{sheng,bhatt}. The additional degeneracy $g_s=4$
around the band center gives rise to a total Chern number $C=4$
for each LL. Thus when each additional LL is occupied, the Hall
conductance increments by $g_{s}\frac {e^2} {h}$. At the
particle-hole symmetric point $E_{F}=0$, corresponding to the
half-filling of the central LL, $\sigma_{xy}=0$ and the total
Chern number of all the occupied states (hole band) must sum up to
zero. Now one can count $\sigma_{xy}$ from this point, and find
that the central LL effectively contributes $\pm
(\frac{g_s}{2})\frac{e^2}{h}$ to $\sigma_{xy}$, when $E_{F}$ is
shifted away from the central LL by adding particles or holes.
This leads to the experimental ``half-integer'' quantization of
$\sigma_{xy}$ in units of $g_s \frac {e^2}{h}$. As will be shown
below, the total zero Chern number for the particle or hole band
resulting from the particle-hole symmetry in the pure system
remains to be true in the presence of disorder.

Near the band edges, each LL carries a total Chern number $C=2$,
and thus the Hall conductance is quantized as
$\sigma_{xy}=kg_s\frac {e^2} h$ with $k$ an integer and $g_{s}=2$
for spin degeneracy only, which is as same as that in the
conventional QHE systems. Remarkably, around $E_{F}=\pm t$, there
are two new critical regions, which separate the unconventional
and conventional QHE states. The extended states in each critical
region carry a large negative total Chern number, e.g., $\sum
C=-72$ for $\phi=\frac {2\pi} {48}$, as shown in Fig.\ 1a. The
existence of the negative Chern number regimes around $E_{F}=\pm
t$ is crucial for understanding the behavior of the Hall
conductance in the whole energy band. When the Fermi energy
$E_{F}$ is increased from the band bottom toward band center
continuously, following a whole sequence of the conventional Hall
plateaus, the negative Chern numbers cause a dramatic reduction
and a sign inversion of $\sigma_{xy}$, so that the unconventional
low Hall plateaus with $\nu=-6,$ $-2,$ $2\cdots$ can reoccur near
the band center. This is in contrast to the QHE on a square
lattice, where stable Hall plateaus can only be observed near the
band edges~\cite{sheng}.
\begin{figure}[tbp]
\par
\includegraphics[width=3.2in]{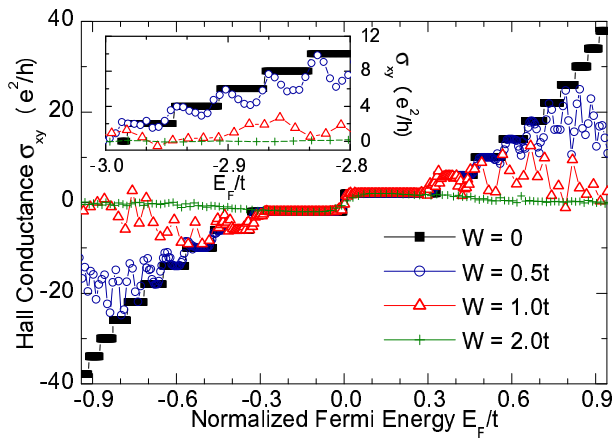}
\caption{Unconventional Hall conductance as a function of electron
Fermi energy near the band center for four different disorder
strengths each averaged over $200$ disorder configurations.
Inset: conventional Hall conductance near the lower
band edge. Here $M=96$ and the sample size is $96\times 48$.}
\vskip-0.3cm
\end{figure}

We have shown that the unconventionally quantized QHE observed in
the experiments can be reproduced in the lattice model, and is due
to the unusual topological property of the energy band. We next
turn to the effect of random disorder on the QHE. In Fig.\ 2, the
Hall conductance around the band center is shown as a function of
$E_{F}$ for magnetic flux $\phi=\frac{2\pi}{96}$ and four
different values of disorder strength $W$. We see that with
increasing $W$, higher Hall plateaus (with larger $|\nu|$) are
destroyed first. At $W=0.5t$, the plateaus with $\nu=\pm 10,\pm 6$
and $\pm 2$ remain well quantized, while at $W=2.0t$ all the
plateaus except for the $\nu=\pm 2$ ones are destroyed. These last
two plateaus will eventually disappear at $W \sim 2.5t$. For
comparison, the QHE near the lower band edge is shown in the
inset, where all plateaus are found to be destroyed at a much
weaker disorder $W=1.0t$. This clearly indicates that, under the
same conditions, the unconventional QHE is much more stable than
the conventional one. This is attributed to the Dirac-fermion-like
linear dispersion relation around the band center, where the
widths of the LL gaps are proportional to $\sqrt B$ instead of
$B$. We also notice that, $\sigma_{xy}$ always vanishes at
$E_{F}=0$ for all $W$, due to the fact that the whole particle or
hole band carries zero total Chern number as a topological invariant
in the disordered system.
\begin{figure}[tbp]
\par
\includegraphics[width=3.2in]{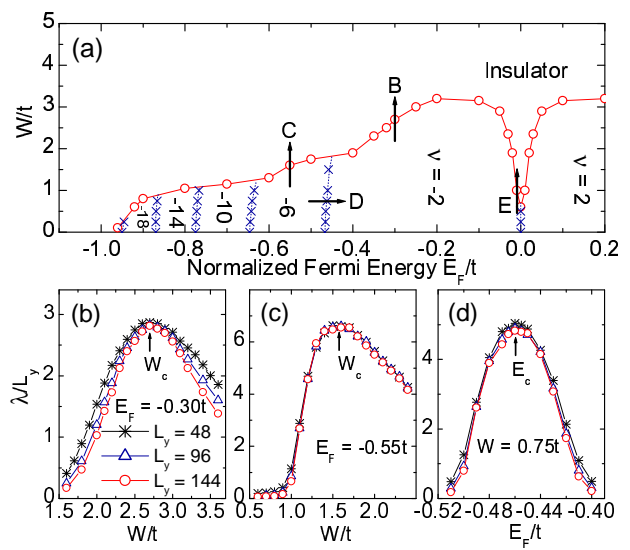}
\caption{(a) Phase diagram for the unconventional QHE regime in
graphene at $M=48$, which is symmetric about $E_F=0$. (b) to (d): 
Normalized localization lengths calculated for three bar widths
 $L_{y}=48$, $96$ and $144$, as the
phase boundary is crossed by the paths indicated by the arrows
$B$, $C$ and $D$ in (a), respectively.} 
\vskip-0.25cm
\end{figure}

We further study the quantum phase transition of the graphene
electron system and establish the phase diagram for the QHE. This
can be done relatively conveniently by calculation of the
finite-size localization length $\lambda$ on an essentially
infinitely long bar of width $L_y$ (length $L_{x}\geq 10^{6}$) by
using the well-established recursive Green's function
approach~\cite{RECUR}. We present the calculated phase diagram in
Fig.\ 3a, for a relatively large magnetic flux
$\phi=\frac{2\pi}{48}$ for clarity, while the topology of the
phase diagram is essentially universal, independent of $\phi$. In
the $W-E_{F}$ plane, different QHE plateaus with $\sigma_{xy}=\nu
\frac {e^2} h$ are separated by extended states, where $\lambda$
grows linearly with increasing bar width $L_y$. With the increase
of $W$, each plateau can be destroyed through a transition $\nu
\rightarrow 0$ to the insulating phase and higher plateaus
disappear first. In Fig.\ 3b-3d, we show examples of calculated
localization length to explain how the phase boundaries in the
phase diagram are determined. In Fig.\ 3b, the normalized
localization length $\lambda/L_y$ for $E_{F}=-0.3t$ and three
sample widths $L_y=48$, $96$ and $144$ is plotted as a function of
$W$, which corresponds to a $\nu=-2\rightarrow 0$ transition as
indicated by the arrow $B$ in the phase diagram. The sample length
$L_{x}$ ranges from $10^{6}$ up to $5\times 10^{6}$, so that the
relative error due to statistical fluctuations in $\lambda$
reduces to about $2\%$. We see clearly that $\lambda/L_y$ is
peaked at $W=W_c\simeq 2.7t$, an indication of an extended
critical point separating the $\nu=-2$ plateau from the outside
insulating phase. Finite-size scaling~\cite{RECUR} confirms that
the localization length at the thermodynamic limit becomes
divergent at $W_c$. This is consistent with a new ``float-up''
picture~\cite{sheng}, where some negative Chern number states are
coming from lower energy, and moving toward the band center with
increasing $W$, which sweep across $E_{F}=-0.3t$ at $W=W_c$,
causing the collapse of the $\nu=-2$ plateau.

Figure 3c shows the normalized localization length as a function
of $W$ at $E_{F}=-0.55t$, corresponding to the path indicated by
the arrow $C$ in the phase diagram Fig.\ 3a. We see that a peak of
localization length occurs at $W=W_c\simeq 1.6t$. Similarly to
Fig.\ 3b, the peak indicates the destruction of the $\nu=-6$ QHE
state and its transition into the insulating phase. However, we
note that here the localization length $\lambda$ is relatively
large, being much greater than the largest $L_{y}$ that is
reachable in our calculations. So $\lambda/L_{y}$ does not
decrease visibly with increasing $L_{y}$, and one cannot rule out
a possibility that the higher plateau to insulator transition
happens in a critical region with a small finite width $\Delta
W_c$ instead of at a critical point $W_c$, where electron states
become delocalized. All the phase boundaries separating the QHE
phases from the insulating phase at strong $W$, indicated by the
solid line with open circles in Fig. 3a, are determined in the
same way.

To determine the phase boundary between different QHE states, the
localization length is calculated as a function of electron Fermi
energy $E_F$ for fixed $W$. As shown in Fig.\ 3d, corresponding to the
path indicated by the arrow $D$ in Fig. 3a, a peak in
$\lambda/L_y$ occurs at $E_{F}=E_c\simeq -0.46t$, which indicates
a critical point separating $\nu=-6$ and $-2$ plateaus. All the
phase boundaries indicated by dotted lines with cross symbols are
determined in the same manner.

\begin{figure}[tbp]
\par
\includegraphics[width=3.2in]{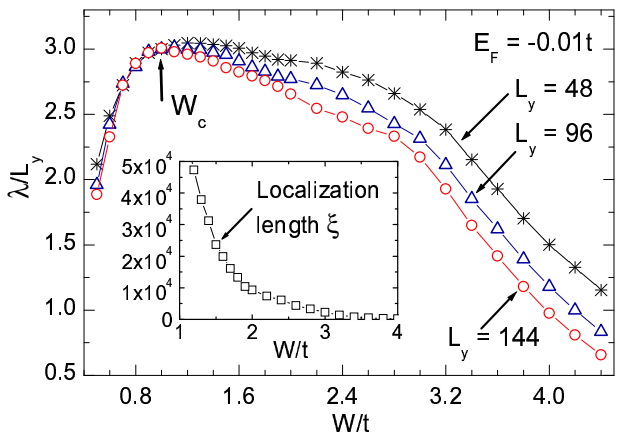}
\caption{The normalized localization length
$\protect\lambda/L_{y}$ for three bar widths $L_{y}=48$, $96$ and
$144$, when the phase boundary is crossed with varying disorder
strength $W$ at $E_{F}=-0.01t$, as indicated by the arrow $E$ in
Fig.\ 3a. Inset shows the localization length
$\protect%
\xi$ at the thermodynamic limit determined through one-parameter
scaling.}
\vskip-0.25cm
\end{figure}
An important feature of the phase diagram is that the $\nu=-2$ and
$2$ plateaus around the band center are no longer connected to
each other in the presence of disorder, separated by a new
insulating phase in between. Corresponding to such a phase
transition, along the path indicated by the arrow $E$ in Fig.\ 3a,
we show $\lambda/L_y$ at $E_{F}=-0.01t$ as a function of $W$ in
Fig.\ 4. Clearly a largely reduced critical disorder strength
$W_c\simeq 1.0t$ is observed in Fig.\ 4. Following the standard
finite-size scaling analysis, we find that all the data in Fig.\ 4
can be well fitted by a one-parameter scaling
relation~\cite{RECUR} $\lambda/L_y=f(L_y/\xi)$ for $L_y=48, 96 $
and $144$. The fitting parameter $\xi(W)$ is the localization
length at the thermodynamic limit. Its value is plotted in the
inset of Fig.\ 4 as a function of $W$, which becomes divergent at
$W_c\simeq 1.0t$. Interestingly, the splitting indicates a
singularity in the resistance near $E_{F}=0$, as an insulating
phase is characterized by a divergent resistance $\rho_{xx}$,
while a plateau state has zero or finite $\rho_{xx}$ (being finite
only at the critical point) at low temperature limit. This feature may
explain the divergent trend in $\rho_{xx}$ and discontinuity in
$\rho_{xy}$ observed near zero gate voltage in the
experiments~\cite{Hall0,Hall1}.

We have also confirmed the above phase boundaries by calculating
the Thouless number~\cite{thouless}, which is proportional to the
longitudinal conductance. In particular, we observed that the
Thouless number as a function of $E_{F}$ at fixed $W$ shows two
peaks near the band center, while a dip occurs at $E_{F}=0$, which
is consistent with the splitting of the $\nu=-2$ to 2 transition
with a new insulating phase emerging near $E_{F}=0$.

In summary, we have numerically investigated the QHE in 2D
graphene based upon a lattice model. The experimentally discovered
unconventional quantization of QHE is reproduced near the band
center, which is understood in terms of the novel distribution of
the topological Chern integers in the energy band. The phase
diagram indicates a new float-up picture, in which the extended
levels move toward band center with increasing disorder strength,
causing higher plateaus to disappear first. The unconventional QHE
plateaus around the band center are found to be much more stable
than the conventional ones near the band edges. A new insulating
phase is predicted to emerge at the band center, between two
$\nu=\pm 2$ QHE states, which is consistent with the
experimentally observed resistance discontinuity near zero gate
voltage.

\textbf{Acknowledgment:} This work is supported by ACS-PRF
41752-AC10, Research Corporation Fund CC5643, the NSF
grant/DMR-0307170 (DNS), a grant from the Robert A. Welch
Foundation under the grant no. E-1146 (LS), and the NSFC grants
10374058 and 90403016 (ZYW).

\end{document}